\documentclass[aps,prl,twocolumn,showpacs]{revtex4}

\usepackage{graphicx} 
\usepackage[usenames,dvipsnames]{xcolor}
\usepackage{tikz}
\usepackage{amsmath}
\usepackage{amssymb}
\usepackage{mathrsfs}
\usepackage{natbib}
\usepackage{color}
\usepackage{psfrag}
\usepackage{wasysym}
\usepackage{float}
\usepackage{epstopdf}
\usepackage[normalem]{ulem}
\usepackage{pgfplots}
\usetikzlibrary{pgfplots.groupplots}
\usepackage{textcomp}
\usepackage{hyperref}
\usepackage{comment}

\pgfplotsset{compat=newest}
\pgfplotsset{yticklabel style={text width=1.3em,align=right}, xticklabel style={text width=1.5em,align=center}}
\pgfplotsset{every axis legend/.append style={anchor=south east}}

\begin{document}
\title{Elastogranular Mechanics: Buckling, Jamming, and Structure Formation}
\author{David J. Schunter Jr., Martin Brandenbourger, Sophia Perriseau, and Douglas P. Holmes}
\affiliation{\footnotesize Mechanical Engineering, Boston University, Boston, MA, 02215, USA}

\date{\today}

\begin{abstract} 

Confinement of a slender body into a granular array induces stress localization in the geometrically nonlinear structure, and jamming, reordering, and vertical dislocation of the surrounding granular medium. By varying the initial packing density of grains and the length of a confined {\em elastica}, we identify the critical length necessary to induce jamming, and demonstrate an intricate coupling between folds that localize along grain boundaries. Above the jamming threshold, the characteristic length of {\em elastica} deformation is shown to scale with the length over which force field fluctuations propagate in a jammed state, suggesting the ordering of the granular array governs the deformation of the slender structure.  However, over confinement of the {\em elastica} will induce a form of stress relaxation in the granular medium by dislocating grains through two distinct mechanisms that depend on the geometry of the confined structure.

\end{abstract}


\maketitle

\begin{figure}[t]
\begin{center}
\vspace{0mm}
\resizebox{1.0\columnwidth}{!} {\includegraphics{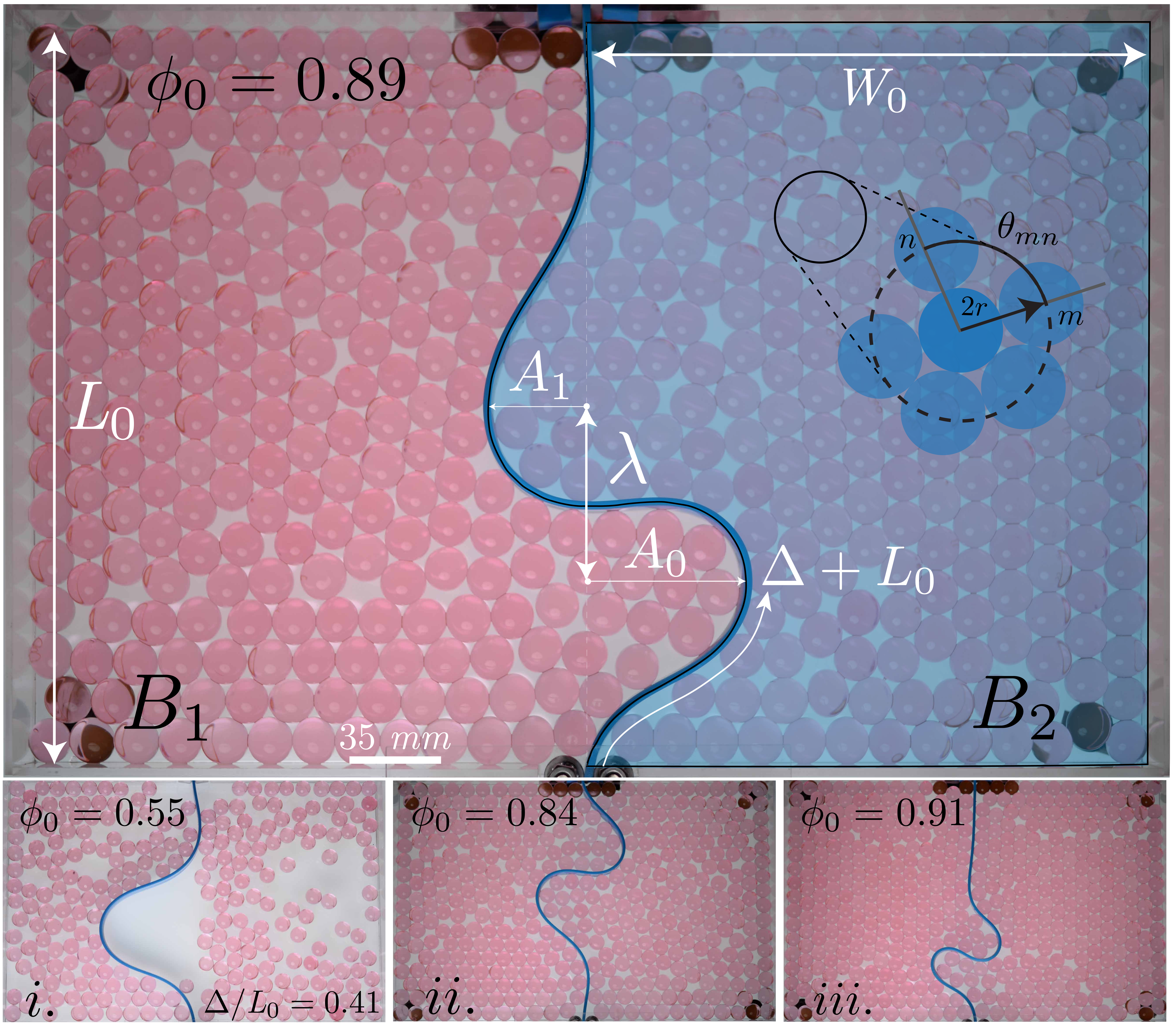}}

\end{center}\vspace{-8mm}
\caption[]{Shape profiles of the elastica as additional arc-length $\Delta$ is injected into a granular array of length $L_0$, width $W_0\,$ and grains of radius ${r}$ over a range of packing fractions: $\phi_0 = 0.55$ (i), $\phi_0 = \phi_j = 0.84$ (ii), $\phi_0 = 0.91$ (iii). 
 \label{fig1}}\vspace{-6mm}
\end{figure}

\indent Consider the growth of an elastic rod within a granular medium. As the rod elongates in a confined space, it will bend to minimize its internal energy ~\cite{audoly2010, capovilla2002}, reordering the surrounding granular material to accommodate higher arc length configurations. At low packing densities, the rod feels little resistance from the grains~\cite{liu1998}, while as the packing density is gradually increased to the point of jamming, the granular matter begins to exert a large, inhomogeneous stress distribution on the elastic rod~\cite{dantu1968, liu1995, coppersmith1996, o'Hern2001}, deforming the geometrically nonlinear structure. It is well-known that slender structures will localize stress in response to a homogeneous stress distribution.  Wrinkled sheets on a fluid substrate exhibit a spontaneous up--down symmetry breaking that tends toward an asymptotic isometry~\cite{pocivavsek2008, holmes2010, diamant2011, rivetti2014, vella2015}, while wrinkled sheets on an elastic substrate exhibit a period doubling instability and subsequent up--down symmetry breaking characterized by a subharmonic mode~\cite{brau2011,brau2013}. Localization of a geometrically nonlinear structure in a discrete medium exerting an inhomogeneous stress distribution is less well understood~\cite{sobral2015, mojdehi2016, gurmessa2017}, although commonly observed in biomechanics. Stresses exerted by soil on a growing root can dictate growth pathways~\cite{bengough2006, kolb2012} and induce chiral, helical buckling~\cite{whiteley1982, oliva2007, silverberg2012}. Further, in dry sand environments, sand vipers can burrow~\cite{young2003}, and desert--dwelling sandfish can swim within a granular bed by propagating an undulatory traveling wave down their rod--like bodies, enabling non-inertial swimming in a frictionless fluid~\cite{maladen2009}.
\indent These coupled, {\em elastogranular} mechanics have generally been considered as local inhomogeneities or studied in systems where the length scale of the rod deformation exceeds by several orders of magnitude the grain size. The question of how granular ordering can influence deformation of a slender body, such as an elastica, has remained open. In this Letter, we describe the connections between jamming, ordering, and stress localization in an elastogranular system through the use of simple scaling arguments, and the observation of the relaxation of stresses within the granular array through the vertical dislocation of grains. These results will help to illuminate the ways slender elastic structures interact with non-homogeneous and fragile media, behavior commonly seen in plant root growth~\cite{silverberg2012}, the piercing of soft tissue~\cite{choumet2012}, and the reinforcement of jammed granular architectures~\cite{fauconneau2016}.

To understand how the discrete, heterogenous behavior of a granular medium couples with a geometrically nonlinear slender structure, we considered the confinement of a planar elastica within a 2D array of frictionless, soft spherical grains. The behavior of the granular material depends on its packing fraction, $\phi_i=\pi N r^2/B_i$, where $r$ is the average grain radius, $N$ is the number of grains, and $B_i$ is the area of the $i^{th}$ side ($i=1,2$). The system exhibits a discontinuity in the material's bulk and shear modulus when $\phi \geq \phi_j$, where $\phi_j=0.84$ (*see Supplemental Material). The geometrically nonlinear behavior of an elastica depends on its bending rigidity per unit width, $Eh^3/12$, where $E$ is Young's elastic modulus, and $h$ is thickness. Buckling and subsequent stress localization in elastic rods are generally characterized by a region of maximum curvature, $\kappa_m \sim A_0/\lambda^2$, where $A_0$ is the primary amplitude, and $\lambda$ is an effective buckling length. Here, we define $\lambda$ as the distance between the two primary maxima of deformation $\textit{A}_0$ and $\textit{A}_1$ (see Fig.~\ref{fig1})~\footnote{This is motivated by the high regularity of quasi-mode two deformations that were observed at full injected arc length. Deformation modes one and three were observed, but less frequently. Higher mode shapes should become more prevalent as the elastica is made thinner or the full injected arc length is increased beyond what was tested.}. We first quantify the elastogranular interactions as the elastica's arc length was increased by $\Delta$ in a quasi--static manner from an initial length $L_0$ for a range of initial packing fractions $\phi_0$ (Fig.~\ref{fig1}$i.$--$iii.$).

\begin{figure}
\begin{center}
\vspace{0mm}
\resizebox{1.0\columnwidth}{!} {\includegraphics{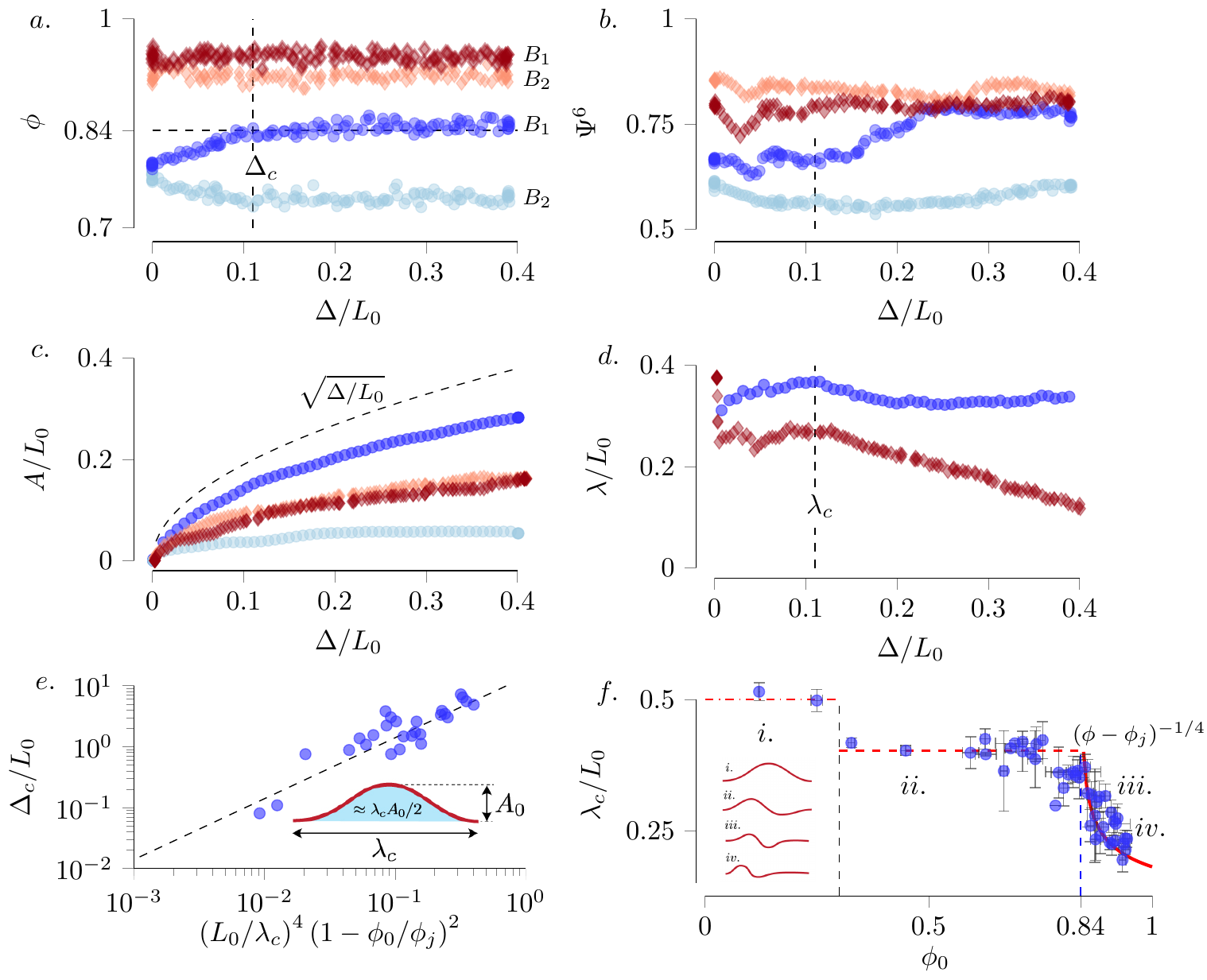}}
\end{center}\vspace{-8mm}
\caption[]{Changes in (a) packing fraction $\phi$, (b) bond-orientation order $\Psi^6$, (c) primary and second amplitude $\textit{A}_0$ and $\textit{A}_1$, and (d) distance between amplitudes $\lambda$, in both pre (blue points) and post-jammed (red diamonds) systems as the elastica is injected $\left(\Delta/L_0\right)$. Two characteristic lengths are seen to emerge:  (e) a critical injected arc-length of elastica $\Delta_c$ necessary to induce jamming in a 2D granular array, and (f) the length of the confinement region $\lambda_c$ in which the elastica will localize curvature.
\label{fig-overtime}}\vspace{-4mm}
\end{figure}

An elastica, clamped at its ends within fixed boundaries, will adopt a cosine--like deflection profile when injected into a low density granular array, {\em i.e.} $\phi_0 \lesssim 0.3$ (see Fig.~\ref{fig1}i.), with its exact shape being governed by an elliptic integral~\cite{bigoni2015}. For larger packing fractions (see Fig.~\ref{fig1}ii.--iii.), we see the typical response of an elastica injected into both non-jammed (blue circles) and jammed states (red diamonds) in Fig.~\ref{fig-overtime}a--d. For initial packing fractions $0.3 \lesssim \phi_0 < \phi_j$, the granular array will provide a very low resistance, and the elastica will assume a quasi mode one deformation shape, with a maximum amplitude that grows as $A_0/L_0 \sim \sqrt{\Delta/L_0}$ (Fig.~\ref{fig-overtime}c -- blue circles). This mode shape breaks the initial left--right symmetry of the 2D array, and the packing fraction on the side containing $A_0$ increases until jamming occurs at a critical elongation $\Delta_c$ (Fig.~\ref{fig-overtime}a -- blue circles). Once the jammed state is reached, the granular packing fraction stays constant as the elastica's arc length is increased. Reordering of the granular array, characterized by the global bond orientation parameter $\Psi^6 = \left |N^{-1} \sum_{m=1}^{N}  N_b^{-1} \sum_{n=1}^{{N}_b} e^{6  i  \theta_{mn}}\right|$, occurs following the onset of jamming (Fig.~\ref{fig-overtime}b -- blue circles), along with a slight drop in normalized distance between maxima, $\lambda/L_0$ (Fig.~\ref{fig-overtime}d -- blue circles). When injected into a solid--like granular array, {\em i.e.} $\phi_0 \geq \phi_j$, the elastica immediately adopts a quasi mode two deformation, with two peaks ($A_0$ and $A_1$) of similar amplitude (Fig.~\ref{fig-overtime}b -- red diamonds). We observe the ordered, granular array become disordered in a region around the two peaks of the elastica to accommodate the excess length, before recrystallizing (Fig.~\ref{fig-overtime}b -- red diamonds). Following this reordering, $\lambda$ values are seen to decrease as additional length is injected into the system. In what follows, we establish a physical model to describe these characteristic elastogranular behaviors. 
	

We first describe the arc length of elastica necessary to induce jamming in an array with $\phi_0 < \phi_j$. The initial half wavelength of the elastica does not vary with $\Delta/L_0$ at low injected arc length for all $\phi_0$. Therefore, we define a characteristic length $\lambda_c$ as the average of $\lambda$ for $0<\Delta/L_0<0.1$.  In a loose granular array, $\lambda_c$ is dictated by the elastica and its boundary conditions, with the primary amplitude growing as $A_0/L_0 \sim \sqrt{\Delta/L_0}$. As injection length becomes larger, the elastica occupies an increasing amount of area within the array. At a critical injection length $\Delta_c$ (dotted line Fig.~\ref{fig-overtime}a), it will decrease the area available to the granular media enough to induce jamming. At small $\Delta$ and $\phi<\phi_j$, the elastica exhibits a primarily mode one shape with $A_0/\lambda \sim \sqrt{\Delta/L}$, which can be approximated as a triangular area of base $\lambda_c$ and height $A_0$ (inset Fig. 2e). As the area on one side of the array is reduced by $\frac{1}{2} \lambda_c^2\sqrt{\Delta/L}$, the packing fraction as a function of $\Delta$ may be written as $\phi= \pi r^2 N/(L_0W_0-\frac{1}{2} \lambda_c^2\sqrt{\Delta/L})$. It follows that by separating the initial packing fraction $\phi_0$, and considering the array at jamming, where $\phi \rightarrow \phi_j$ and $\Delta \rightarrow \Delta_c$, we can arrive at a critical length of elastica needed to jam an array of 2D, frictionless, spherical grains, {\em i.e.} an effective {\em elastogranular} length, 
\begin{equation}
\label{Dc}
\frac{\Delta_c}{L_0} \sim \left(\frac{L_0}{\lambda_c}\right)^4\left[1-\frac{\phi_0}{\phi_j}\right]^2, 
\end{equation}
where the wavelength $\lambda_c$ is independent of $\phi$ and $\Delta$ for $\phi < \phi_j$ (Fig.~\ref{fig-overtime}d). Equation~\ref{Dc} is plotted in Fig.~\ref{fig-overtime}e along with arrays that jammed following the injection of $\Delta_c$, and there is a strong agreement with the prediction. 




Beyond the jamming threshold, the elastica always localizes deformation over a finite length smaller than $L_0$ (Fig.~\ref{fig-overtime}f), similar to its behavior on a homogenous elastic foundation~\cite{hetenyi1946, pocivavsek2008, holmes2010, diamant2011, rivetti2014}. Characteristic lengths arising from the competition between the internal energies of the elastic beam and supporting substrate provide accurate models of these continuous systems. An elastica coupled with a discrete medium necessitates a different approach. Studies on the response of grains to point forces show a zone of high fluctuation in the granular stress field that will delineate a region of localization~\cite{o'Hern2013, wyart2005, vanhecke2009}. If $\lambda_c$ scales with the hypothetical length scale governing this fluctuation area (*see Supplemental Material), then we would expect, 
\begin{equation}
\label{lstar}
\lambda_c \sim \ell_c\sim\frac{1}{(\phi-\phi_j)^{1/4}}.
\end{equation}
Comparison of the diverging initial wavelength $\lambda_c/L_0$ with Eq.~\ref{lstar} beyond the jamming point shows a good agreement (Fig.~\ref{fig-overtime}f), given the expected scatter in local packing fraction~\cite{lerner2014}. A look at the experimentally observed elastica shapes at different values of $\phi_0$ shows the characteristic deformation shape being confined to a smaller length along the elastic arc length (Fig.~\ref{fig-overtime}f, i--iv). The scaling of the characteristic length of deformation of the elastica with the length over which force field fluctuations propagate in a jammed state, suggests that the interaction between grains governs the deformation of the slender structure. 

\begin{figure}
\begin{center}
\vspace{0mm}
\resizebox{1.0\columnwidth}{!} {\includegraphics{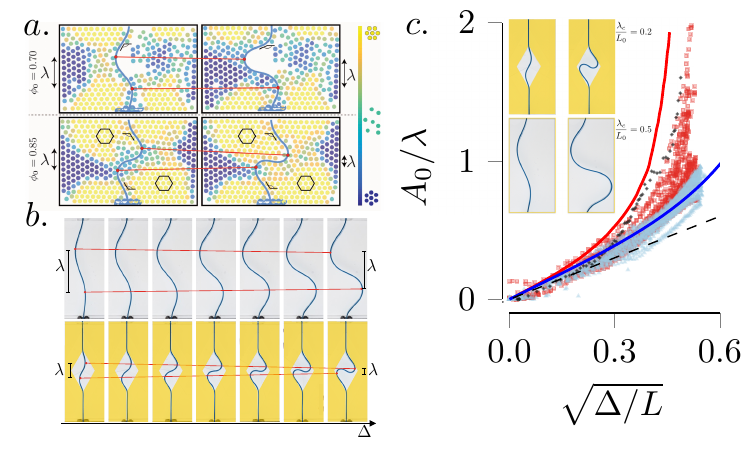}}
\end{center}\vspace{-8mm}
\caption[]{(a) For $\phi_0 < \phi_\textit{j}$, the elastica deforms to one side of the enclosure, modifying its crystal structure. At $\phi_0 \geq \phi_\textit{j}$, a symmetric deformation of the elastica grows along the existing crystal structure. (b) Model experiments demonstrate that the crystal structure acts as a rigid boundary confining the elastica, shown as a lozenge of characteristic length $\lambda_c$. (c) A plot of $A_0/\lambda$ as a function of $\sqrt{\Delta/L}$. When $\phi < \phi_\textit{j}$ (blue triangles), the deformation follows the shape of a free elastica (blue solid line), and when $\phi > \phi_\textit{j}$ (red squares), it is confined by an upper bound corresponding to the lozenge--like crystal structure, verified both experimentally (black circles) and numerically (red line). \label{fig3}}\vspace{-4mm}
\end{figure}



To understand how the local packing and order of the granular array influences the shape of the confined elastica, we compare the elastica shape and granular ordering of two typical experiments ($\phi_0=$ 0.70 and $\phi_0=$ 0.85) in Fig.~\ref{fig3}a, where each grain is colored by a measure of its local bond orientation number, $\psi_m^6 = N_b^{-1} \sum_{n=1}^{{N}_b} e^{6  i  \theta_{mn}}$. In the non-jammed array, the grains move freely to accommodate the growing amplitudes of the elastica, while the jammed grains have a significant influence on the shape of the elastica. In Fig.~\ref{fig3}a, when ($\phi_0=0.85$), regions near the fixed end of the elastica are surrounded by hexagonally packed grains~\cite{torquato2001}. The deformation of the elastica is bound by these regions, forming an angle of sixty degrees with the horizontal. Disordered grains near $A_0$ and $A_1$ adopt the same hexagonal orientation as the elastica elongates, the half wavelength $\lambda$ decreases, and the elastica tends towards an antisymmetric, overlapping fold -- a shape expected for large folds on fluid interfaces, but not commonly observed~\cite{demery2014}. Notably, $\lambda$ remains constant for $\phi < \phi_j$, yet decreases for $\phi \geq \phi_j$. 

These competing behaviors can be illustrated using two simple experimental models: an elastica can either freely elongate in mode two, or be confined by rigid walls representing the restriction imposed by the grains (Fig.~\ref{fig3}b). These models should represent the bounds of the elastogranular behavior, and we confirm this by plotting $A_0/\lambda$ as a function of $\sqrt{\Delta/L}$ over a large range of initial packing fraction ($0.1<\phi_0<0.97$) in Fig.~\ref{fig3}c. For all $\phi_0$ and $\Delta/L < 0.3$, the normalized amplitude scales linearly with $\sqrt{\Delta/L}$ (Fig.~\ref{fig3}c -- dashed black line). At larger confined lengths, the ratio of amplitude to wavelength strongly depends on whether the elastica is injected into a loose (blue triangles) or jammed (red squares) granular state. Within a loose granular array $A_0/\lambda$ follows the shape of the antisymmetric, nonlinear elastica~\cite{bigoni2015} -- {\em i.e} a freely injected elastica (Fig.~\ref{fig3}c -- solid blue line)(*see Supplemental Material). The ratio of $A_0/\lambda$ rapidly diverges from the classical behavior when the elastica elongates within a jammed array, as the crystal structure of the grains geometrically confines the elastica. The shape of the confinement will be dictated by the packing fraction, and the local orientational order around the elastica. From Eq.~\ref{lstar}, we expect the length scale of this confinement to be governed by $(\phi-\phi_j)^{-1/4}$, with the highest confinement occurring when the crystal structure near the primary maxima $A_0$ and $A_1$ forms an angle of sixty degrees with the horizon. Confinement within this space represents an upper bound on the diverging ratio of $A_0/\lambda$, as demonstrated by experiments using rigid walls (image sequence Fig.~\ref{fig3}b and black points Fig.~\ref{fig3}c) and by numerically solving the elastica equation within lozenge--shaped voids (red line Fig.~\ref{fig3}c, *see Supplemental Material).
	
\begin{figure}
\begin{center}
\vspace{0mm}
\resizebox{1.\columnwidth}{!}{\includegraphics{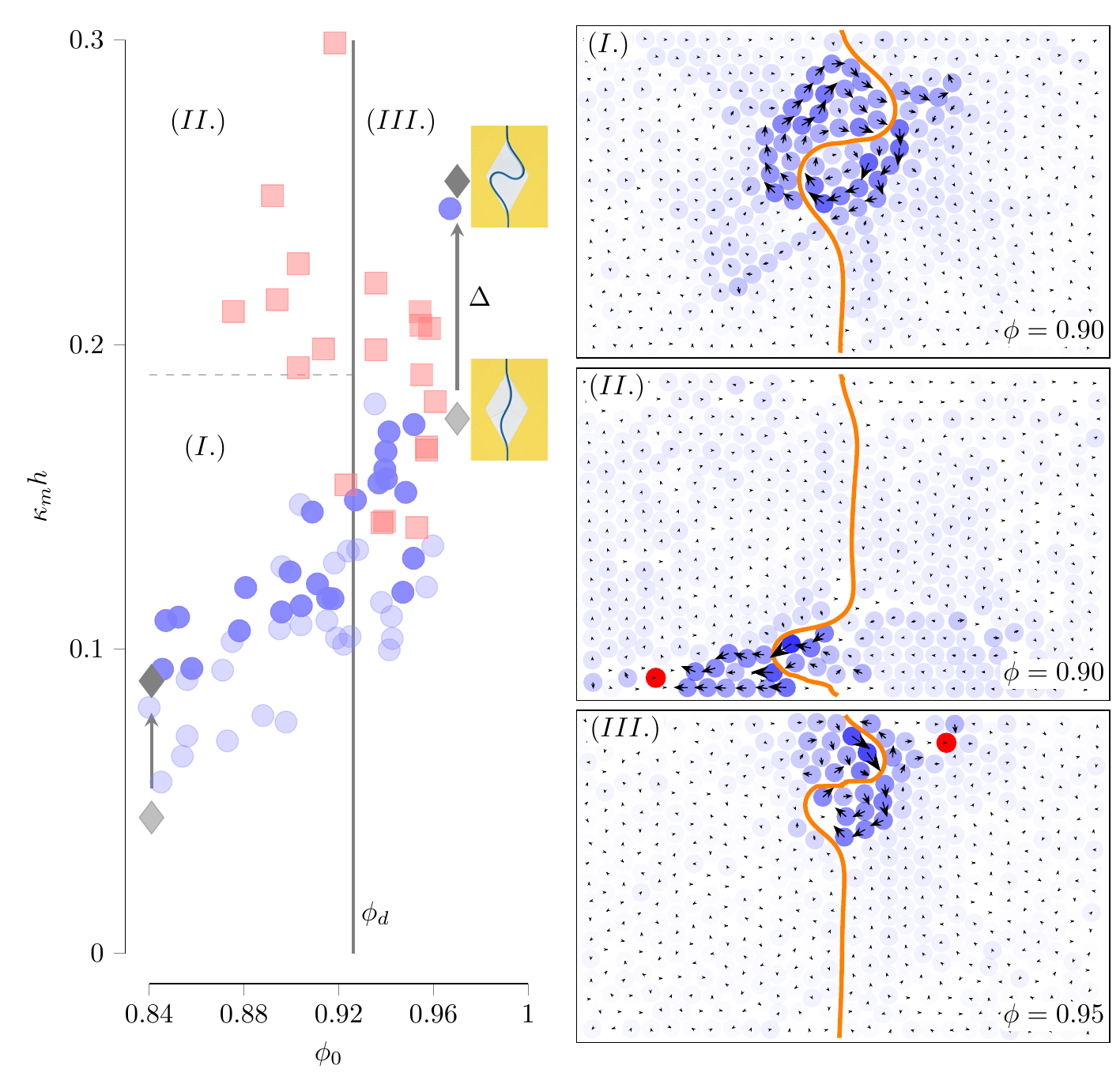}}
\end{center}\vspace{-8mm}
\caption[]{(a) Elastica maximum curvature multiplied by its thickness $\kappa h$ as a function of the initial packing fraction $\phi_0$. The light and dark blue points describe $\kappa h$ for an injected arclength $\Delta/L_0=$ 0.1 and 0.41 respectively. The red squares correspond to $\kappa h$ preceding a vertical dislocation. The light and dark gray diamonds correspond to measurements from experiments with rigid boundaries (see associated images). System behavior is grouped into three distinct regions $\left(I,II,III\right)$. (b) Experimental examples of the vector fields of the bead displacement associated with each region. The bead opacity corresponds to the norm of their vector displacement. \label{fig4}}\vspace{-4mm}
\end{figure}	

It appears from Fig.~\ref{fig3}c that the elastica governs the elastogranular behavior when $\phi < \phi_j$, while the granular array governs the behavior when $\phi \geq \phi_j$, however this trend breaks down at high packing fractions or in rare cases where we observe highly localized elastica folds. At large enough confinement, the granular monolayer can yield by vertically dislocating a bead~\cite{tordesillas2014}. In Fig.~\ref{fig4}, we plot the maximum curvature of the elastica normalized by its thickness, $\kappa h$, as a function of the grains packing fraction $\phi$ for a short and a long injected arc length ($\Delta/L_0$ = 0.11 and 0.41 for the light and dark blue circles, respectively), and indicate the curvature at which a bead was dislocated (red squares). We note three regions in this plot. In region $I$, we observe an equilibrium elastica shape, and no grain dislocations. Tracking the displacement vectors of each bead for a characteristic experiment in this region, we see that a high number of beads close to the primary maxima tend to rotate around the deforming beam (Fig.~\ref{fig4}$I$). The initial arrangement of the beads can force the elastica to localize with a high curvature and because granular motion tends to focus in a given direction, the highly curved beam can act like a point force within the array (Fig.~\ref{fig4}$II$). Therefore, at the same packing fraction, we can sometimes observe more highly confined elastica shapes containing folds of high curvature, which can induce a dislocation within the granular array (Fig.~\ref{fig4}$II$). Finally, beyond a critical packing fraction, dislocation appears to be independent of $\kappa_m h$ (Fig.~\ref{fig4}$III$). To understand the role of packing fraction on dislocation, we homogeneously reduced the area occupied by a monolayer of beads absent of an elastica, and measured $\phi$ at the first dislocation event. A small perturbation beyond a critical packing fraction of $\phi_d=$0.926 (black vertical line) generates a dislocation, suggesting that the elastogranular dislocations correspond to the packing limit of these soft beads~\footnote{The experimentally determined $\phi_d$ is larger than the hexagonal packing limit of $\phi_h=0.907$, and is likely due to the deformability and slight polydispersity of the beads.}. Here again we observe a similar granular displacement field as seen in region $I$, though confined to a smaller region as expected from equation~\ref{lstar}.

The wealth of elastogranular behaviors observed here indicate an intricate coupling between geometrically nonlinear slender bodies and heterogenous, fragile matter. The confinement and deformation of the slender structure is highly dependent on the proximity of the granular array to the jamming point, yet the competition between the structure's elastic energy and the granular matter's local order gives rise to a variety of elastogranular behaviors (notably antisymmetric/overlapping folds and a deformation length scale proportional to packing fraction) that can be observed across a range of packing fractions and confined lengths. These results will bring new insight into the behavior of deformable structures within granular matter, colloidal systems, and soft gels, and will be relevant to modeling root growth and developing smart, steerable needles. \vspace{-5mm}

\section{Acknowledgements}
DPH and DJSJ acknowledge the financial support from NSF CMMI -- CAREER through Mechanics of Materials and Structures (\#1454153). MB acknowledges the financial support from BAEF. DPH thanks Frederic Lechenault and Evelyne Kolb for helpful discussions and inspiration. We also thank Osiagwe Osman for performing preliminary experiments, Michele Curatolo for his help on COMSOL software and Abdikhalaq Bade for implementing the particle image velocimetry to track the granular motion. 

\bibliography{elastica_grow2.bib}  

\clearpage

\section{Supplemental Material}

\begin{center}	\textbf{Materials \& Methods	}\end{center}

	For our experiments, a polyvinyl siloxane (Zhermack) elastica with Young's modulus \textit{E} = 0.8 \textit{MPa}, thickness \textit{h} = 3.175 \textit{mm}, width \textit{b} =  2.225 \textit{mm}, initial arc-length $\textit{L}_0$ = 279.4 \textit{mm} is positioned along the median axis of an initially empty rectangular acrylic box (McMaster-Carr) with dimensions: 438.15 $\times{}$ 279.4 $\times{}$ 34.93 $\textit{mm}$. The beam is secured via boundary conditions of the clamped-roller type [see Fig. 1], resulting in two equal sized areas $\textit{B}_1$ and $\textit{B}_2$. Equal numbers of hydrogel beads (MagicWaterBeads), with quantities ranging from $0 \leq N \leq 238$, are then introduced into $\textit{B}_1$ and $\textit{B}_2$, such that the total number of grains, $\Sigma$ = 2\textit{N}. The box was placed on a 45.742 $\times{}$ 60.96 ${mm}$ LED light panel (Porta-Trace/Gagne). The elastica was quasi-statically injected into the experimental enclosure via a linear actuator (Zaber), with a maximum injected length $\Delta_M = 114.3\  \textit{mm}$. A custom mount was fabricated to ensure vertical orientation of the beam during entry, with no angle of influence. Experiments were documented with time lapse photos at at intervals of 5 \textit{FPS} (Canon D-610 digital camera; Nikon 55 \textit{mm} Micro-NIKKOR lens).
The injection of the elastic membrane within lozenge--shaped voids was characterized via the software COMSOL. We used the classic theory of an elastic rod and step functions to simulate the rigid walls. 
\begin{center}	\textbf{Characterizing the Granular Medium}	\end{center}

	The granular medium is composed of initially dry hydrogel beads (MagicWaterBeads), swollen to their maximum radius of $8.9\pm.4\ \textit{mm}$ in a soap/water solution (1:300). We assume a frictionless system of soft, mono-disperse, spherical grains. These assumptions were validated by comparing experimental values of the average contact number per particle $\textit{Z}$, and the global bond-orientation order parameter $\Psi^6$, with those cited in the literature. In calculating $\textit{Z}$ and $\Psi^6$, we disregard grains in contact with the finite boundaries of the enclosure, moving in a distance of $2{r}$ from all sides. Fig. 5a shows that $\phi_j \approx .84$ occurs when $\textit{Z} = 4$, the isostatic minimum of (on average) 4 contacts per grain that characterizes a jammed, frictionless granular system. Fig. 5b compares the experimental results of evolving bond-orientation order with control experiments conducted with dry glass marbles. Experiment and control data are in good agreement, the vertical shift in experimental measurements resulting from the unavoidable effects of capillary bridging arising in the use of hydrogel beads. A shift in the experimental data below the control data would be an indication of friction in the system.

\begin{figure}
\begin{center}
\vspace{0mm}
\resizebox{1.0\columnwidth}{!} {\includegraphics{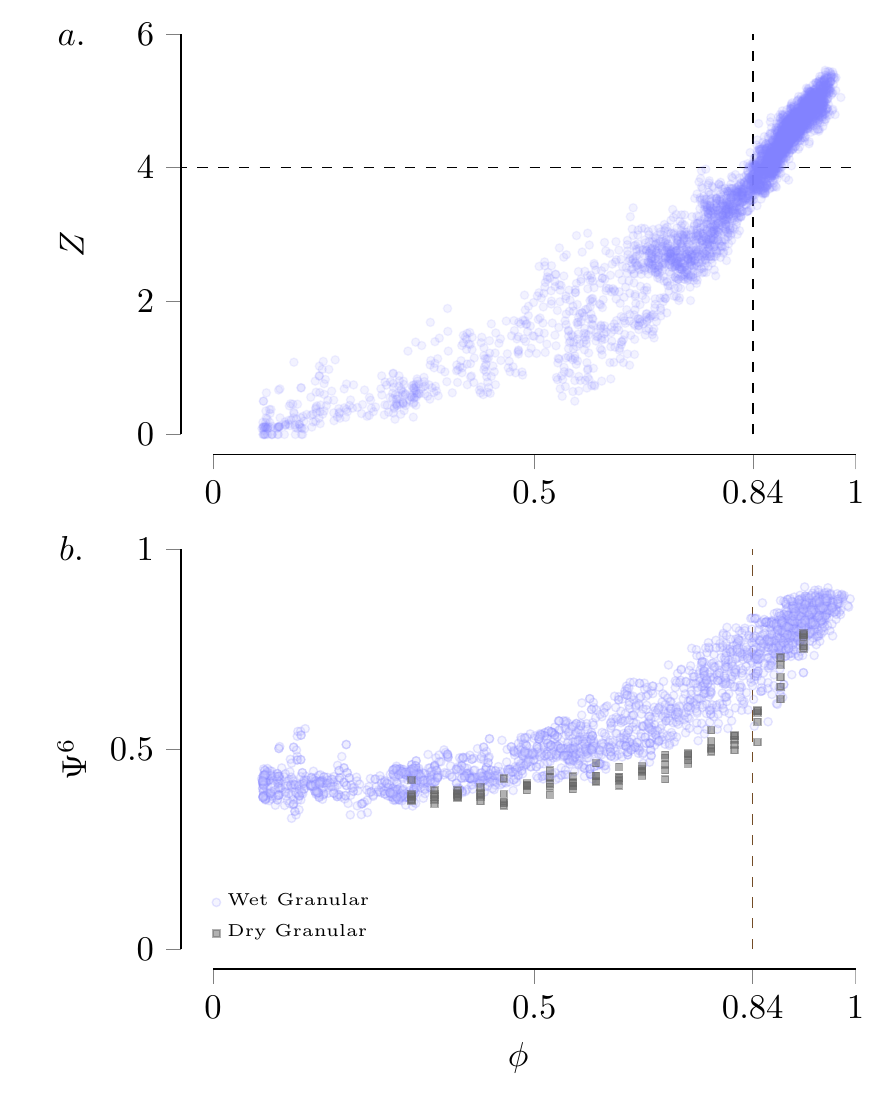}}
\end{center}\vspace{-6mm}
\caption[]{Experimental values for (a) average contact number and (b) global bond orientation order over a range of $\phi$ values.   \label{fig-z-psi}}\vspace{-4mm}
\end{figure}
	
\vspace{15mm}	
	
\begin{center}	\textbf{``Growth'' of a Free Elastica}	\end{center}

\begin{figure}
\begin{center}
\vspace{0mm}
\resizebox{0.5\columnwidth}{!} {\includegraphics{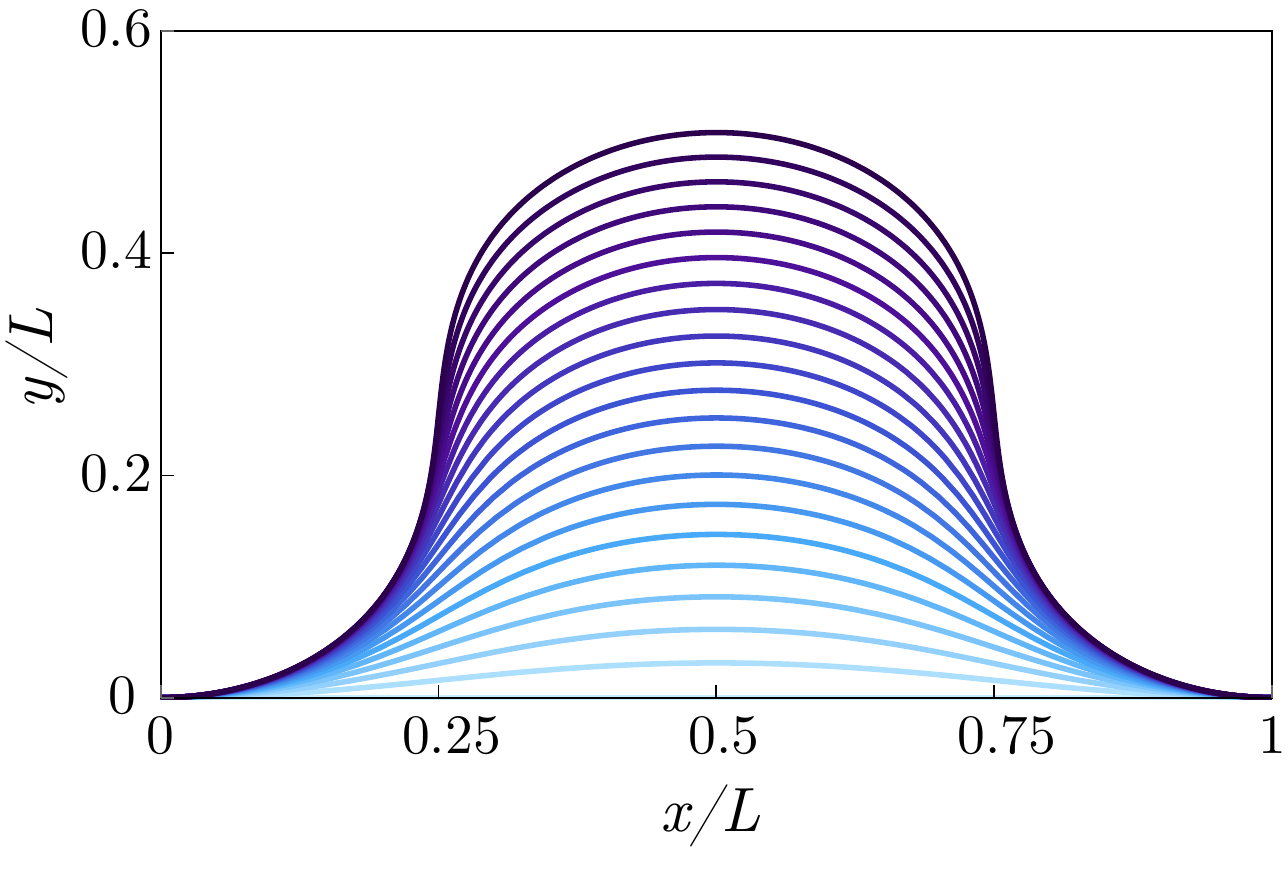}}\resizebox{0.6\columnwidth}{!} {\includegraphics{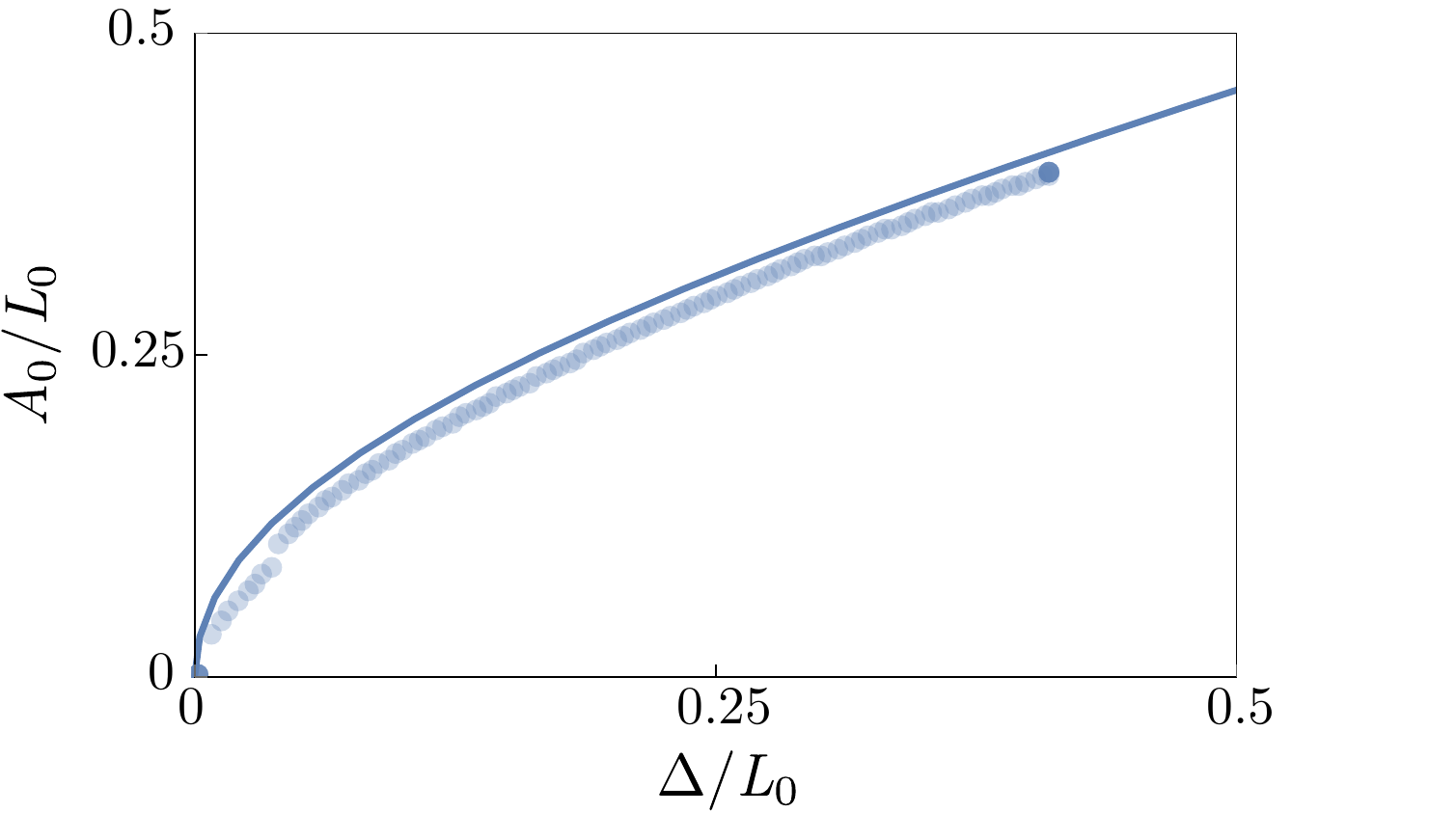}}
\end{center}\vspace{-6mm}
\caption[]{Parametric plots of elastica's of different arc lengths from equations~\ref{eq-xg}, and a plot of $A_0/L_0$ vs. $\Delta/L_0$ to compare the model with our experimental results of the freely injected case. \label{fig-growArch}}\vspace{-4mm}
\end{figure}

The equilibrium equation of a planar elastica is given by
\begin{equation}
\label{eq-elastica}
\theta''(s)+\Lambda^2 \sin \theta(s) =0,
\end{equation}
where $s$ is the arc length that parameterizes the curve, $\theta$ is the angle that the tangent vector at a given point $s$ makes with the horizon, and $\Lambda^2=P/B$ is the ratio of the applied load $P$ to the bending rigidity $B$ of the beam~\cite{singh2017}. The post--buckling shape of clamped--clamped elastica is then determined by the parametric equations~\cite{bigoni2015}
\begin{subequations}
\begin{align}
\label{eq-xs}
x(s) &= \frac{2}{\Lambda(k)}\mathcal{E}[\text{am}[s \Lambda(k)|k]|k]-s, \\
y(s) &= \frac{2}{\Lambda(k)}k(1-\text{cn}[s\Lambda(k)|k]),
\end{align}
\end{subequations}
where $k=\sin{\frac{\alpha}{2}}$, $\alpha$ is the angle of rotation at the inflection point at $s=L/4$, {\em i.e.} symmetry allows the analysis to focus on only quarter of the rod length, $\Lambda(k)=2(m+1)\mathcal{K}[k]$ which for a mode one deformation $m=1$, $\mathcal{K}[\cdot]$ is the complete elliptic integral of the first kind, $\mathcal{E}[\cdot|\cdot]$ is the incomplete elliptic integral of the second kind, $\text{am}[\cdot|\cdot]$ is the amplitude for Jacobi elliptic functions, and $\text{cn}[\cdot | \cdot]$ is the Jacobi cn elliptic function. Increasing $\alpha$ will increase the amplitude of the elastica while conserving the elastica's arc length. For an elastica that is elongating between two fixed ends, we can multiply the parametric equations by a scalar $\Gamma$ representing the increment in growth, 
\begin{subequations}
\begin{align}
\label{eq-xg}
x_g(s) &= \Gamma x(s), \\
y_g(s) &= \Gamma y(s).
\end{align}
\end{subequations}
For a given value of $\Gamma$, we determine the angle $\alpha$ that matches the boundary condition at the end of the elastica, {\em i.e.} $x(1)=1$, by numerical root finding using Newton methods (Fig.~\ref{fig-growArch}--left). To determine the injected length $\Delta$, we numerically integrate the parametric curves given by equations~\ref{eq-xg} for a range of $\Gamma$ to calculate their arc length $L$, and compared this with our experimentally measured values of the freely injected elastica (Fig.~\ref{fig-growArch}--right). For the injection of a free elastica in the absence of grains $\lambda=L_0/2$, and $A_0=y_g(1/2)$, and the resulting curve $A_0/\lambda$ vs. $\sqrt{\Delta/L}$ is plotted in Fig.~\ref{fig3}c. Scatter of the data around this curve is expected due to the presence of grains which can either decrease $\lambda$ or localize the bending in the elastica, thereby increasing $A_0$.

\begin{figure}
\begin{center}
\vspace{0mm}
\resizebox{1.0\columnwidth}{!} {\includegraphics{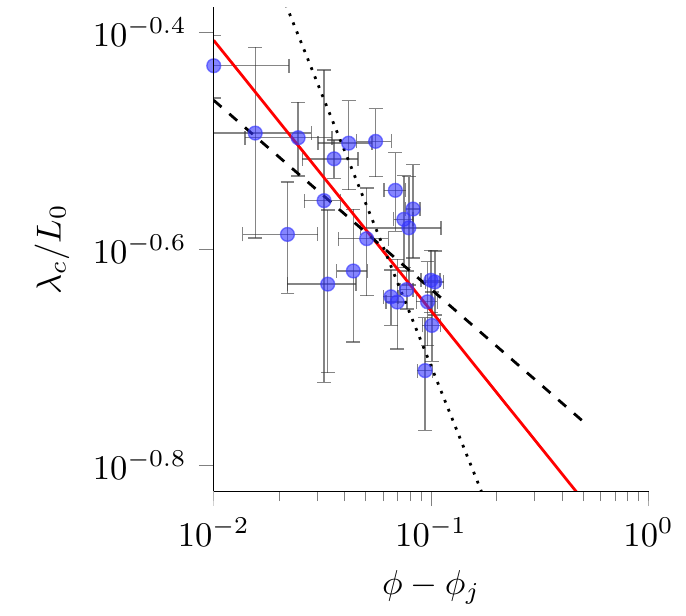}}
\end{center}\vspace{-6mm}
\caption[]{Logarithmic plot of the characteristic length $\lambda_c/L_0$ as a function of $\phi-\phi_j$. The black dotted line corresponds to the best power law fit. The red line and the black dashed line correspond respectively to fits from the lengths $\ell_c$ and $\ell^*$. \label{fig-lambdaclog}}\vspace{-4mm}
\end{figure}

\vspace{24mm}	
\begin{center}	\textbf{Characteristic Length $\lambda_c$	}\end{center}
As stated in this Letter, the characteristic length describing the region of force fluctuation in granular media is still subject to discussion~\cite{baumgarten2017}. $\ell_c$ (see Eq. 2) and $\ell^* \sim (\phi-\phi_j)^{-0.5}$ represent two commonly agreed upon scaling lengths used to describe phenomena in granular materials. Although the current consensus is to describe regions of force fluctuation with $\ell_c$~\cite{lerner2014}, we fit our measurements in Fig.~\ref{fig-lambdaclog} with both scaling lengths, as well as a power law $(\phi-\phi_j)^\beta$. The best power law fit is found for $\beta=$ 0.17. From these observations, we determine that the length scale best describing our measurements is $\ell_c$.

\end{document}